\documentstyle[12pt]{article}
\begin{document}
\pagestyle{empty}
\hspace*{12.4cm}IU-MSTP/28 \\
\hspace*{13cm}hep-th/9802188 \\
\hspace*{13cm}February, 1998
\begin{center}
 {\Large\bf Exact Operator Solution of $A_2$-Toda Field Theory}
\end{center}

\vspace*{1cm}
\def\thefootnote{\fnsymbol{footnote}}
\begin{center}{\sc Takanori Fujiwara,}$^1$
{\sc Hiroshi Igarashi}
and  {\sc Yoshio Takimoto}
\end{center}
\vspace*{0.2cm}
\begin{center}
{\em $\ ^{1}$ Department of Physics, Ibaraki University,
Mito 310, Japan}\\
{\em Graduate School of Science and Engineering,
Ibaraki University, Mito 310, Japan}\\
\end{center}
\vfill
\begin{center}
{\large\sc Abstract}
\end{center}
\noindent
Quantum $A_2$-Toda field theory in two dimensions is investigated based on the 
method of quantizing canonical free field. Toda exponential 
operators associated with the fundamental weights are constructed to the 
fourth order in the cosmological constant. This leads us to a conjecture 
for the exact operator solution.

\vskip .3cm
\noindent

\noindent

\newpage
\pagestyle{plain}

\noindent
It is well-known that Toda field theories in two 
dimensions admit exact classical solutions \cite{ls}. Recently, they 
have attracted renewed interests as conformal field theories 
incorporating $W$-symmetry \cite{wsymm,bilal,mansfield}. 
The simplest Toda theory is the Liouville theory, for which extensive 
canonical approaches have been developed so far 
\cite{gn,bcgt,dhj}. This includes the exact operator solution
\cite{ow,cgs,kn,fit96}. Though similar development have been achieved 
also for Toda theories \cite{mansf82,fl,bilal}, complete operator 
solution seems to be still lacking. 

In this note we investigate exact operator solution of $A_2$-Toda 
field theory by the method of quantizing canonical free field developed 
for Liouville theory \cite{gn,bcgt,dhj}. We shall closely follow the 
arguments due to Otto and Weigt \cite{ow} in constructing the operator 
solution for Liouville theory. It should be noted, however, 
that the extension of their method to Toda theories is not so 
obvious since the latter contain more than one screening charges
which are not mutually commuting in general. Furthermore, Toda 
exponential operators can be parametrized by an arbitrary vector in the 
root space. These brings about an operator ordering problem in 
expanding the Toda exponentials in terms of the screening charges.

Such complications may be resolved by considering special operators 
associated with the fundamental weights. We find that only commuting 
screening charges appear in the expansions of these operators.
Their quantum expressions can be determined 
order by order in the cosmological constant from the requirement of 
locality \cite{bcgt,ow}. General exponential operators as well as the 
Toda field itself can be obtained from them. Conformality and locality 
alone determine the exact solution up to a constant related to the 
arbitrariness of the cosmological constant. This ambiguity is resolved 
by imposing the field equation. The canonicity of the transformation 
from the Toda field to the free field at the quantum level can be 
established directly by examining the canonical commutation relations. 

Let us consider $A_2$-Toda field theory described by the classical action
\begin{equation}
  \label{EQ1}
  S=\frac{1}{\gamma^2}\int_{-\infty}^{+\infty} d\tau
  \int_0^{2\pi}d\sigma\Bigl(\frac{1}{2}\partial_\alpha\varphi\cdot
  \partial^\alpha\varphi
  -\mu^2\sum_{a=1,2}{\rm e}^{\alpha^a\cdot\varphi}\Bigr) ~,
\end{equation}
where $\varphi$ is a two-component field and $\alpha^a$ 
$(a=1,2)$ stand for the simple roots normalized to $(\alpha^a)^2=2$. 
The coupling constant $\gamma$ may be fixed by the conformal 
invariance in the presence of matter couplings. 
We assume that $\varphi$ is subject to periodic boundary condition 
$\varphi(\tau,\sigma+2\pi)=\varphi(\tau,\sigma)$. 
It is well-known that the theory admits exact classical solution. We 
write it in the following form 
\begin{equation}
  \label{EQ2}
  {\rm e}^{\lambda^a\cdot\varphi}
  =\frac{{\rm e}^{\lambda^a\cdot\psi}}{1+\displaystyle{\frac{\mu^2}{4}
    A_aB_a+\biggl(\frac{\mu^2}{4}\biggr)^2}A_a\circ A_{\bar a}B_a
    \circ B_{\bar a}} ~,
\end{equation}
where $\lambda^a$ $(a=1,2)$ are the fundamental weights 
satisfying $\lambda^a\cdot\alpha^b=\delta^{ab}$ 
and $\psi(x)=\psi_+(x^+)+\psi_-(x^-)$ is the canonical free field with the 
normal mode expansion
\begin{eqnarray}
  \label{EQ3}
  \psi_\pm(x^\pm)=\frac{\gamma}{2}Q+\frac{\gamma}{4\pi}Px^\pm
  +\frac{i\gamma}{\sqrt{4\pi}}\sum_{n\neq0}\frac{1}{n}a^{(\pm)}_n
  {\rm e}^{-inx^\pm} ~.
\end{eqnarray}
The light-cone coordinates are defined by $x^\pm=\tau\pm\sigma$, and  
we have introduced 
\begin{eqnarray}
  \label{EQ4}
  A_a(x^+)&=&C_{\alpha^a}\int_0^{2\pi}dy^+{\cal E}_{\alpha^a}(x^+-y^+)
  {\rm e}^{\alpha^a\cdot\psi_+(y^+)} ~,\nonumber\\
  B_a(x^-)&=&C_{\alpha^a}\int_0^{2\pi}dy^-{\cal E}_{\alpha^a}(x^--y^-)
  {\rm e}^{\alpha^a\cdot\psi_-(y^-)} ~,\nonumber\\
  A_a\circ A_{\bar a}(x^+)&=&C_{\alpha^a+\alpha^{\bar a}}C_{\alpha^{\bar a}}
  \int_0^{2\pi}dy^+dz^+{\cal E}_{\alpha^a+\alpha^{\bar a}}(x^+-y^+)
  {\cal E}_{\alpha^{\bar a}}(y^+-z^+){\rm e}^{\alpha^a\cdot\psi_+(y^+)
    +\alpha^{\bar a}\cdot\psi_+(z^+)} ~,
  \nonumber\\
  B_a\circ B_{\bar a}(x^-)&=&C_{\alpha^a+\alpha^{\bar a}}C_{\alpha^{\bar a}}
  \int_0^{2\pi}dy^-dz^-{\cal E}_{\alpha^a+\alpha^{\bar a}}(x^--y^-)
  {\cal E}_{\alpha^{\bar a}}(y^--z^-){\rm e}^{\alpha^a\cdot\psi_-(y^-)+
    \alpha^{\bar a}\cdot\psi_-(z^-)} ~,
\end{eqnarray}
where $\displaystyle{C_\beta\equiv\sinh\frac{\gamma}{4}
\beta\cdot P}$ and ${\cal E}_\beta(x)\equiv
\displaystyle{\exp\frac{\gamma}{4}\beta\cdot 
P\epsilon(x)}$ with $\epsilon(x)$ being the stair-step function defined 
by $\epsilon(x)={\rm sign}(x)$ for $-2\pi<x<2\pi$ and $\epsilon(x+2\pi)
=\epsilon(x)+2$. Furthermore, the convention $\bar 1(\bar 2)=2(1)$ has 
been employed  for the indices. These functions satisfy the periodicity 
$A_a(x^++2\pi)={\rm e}^{\frac{\gamma}{2}\alpha^a\cdot P}A_a(x^+)$, 
$A_a\circ A_{\bar a}(x^++2\pi)={\rm e}^{\frac{\gamma}{2}(\alpha^a+
\alpha^{\bar a})\cdot P}A_a\circ A_{\bar a}(x^+)$, etc. 
One can easily realize that (\ref{EQ2}) is a generalization of the classical 
solution for Liouville theory. In the $A_2$-Toda theory we have four 
screening charges  $A_aB_a$ and $A_a\circ A_{\bar a}B_a\circ B_{\bar a}$ 
($a=1,2$), while there is only one in the Liouville case. 

The key property that enable us to solve the theory exactly is that 
(\ref{EQ2}) defines a canonical mapping from the interacting Toda field 
to the free field. By extending the analysis of ref. \cite{ow,kn} for 
Liouville theory, one can show the fundamental Poisson 
brackets for $\varphi$ and the conjugate momentum $\pi_\varphi=
\displaystyle{\frac{1}{\gamma^2}\dot\varphi}$ by assuming the Poisson 
brackets for the canonical free fields
\begin{eqnarray}
  \label{EQ5}
  \{\psi_k(\tau,\sigma),\dot\psi_l(\tau,\sigma')\}
  =\gamma^2\delta_{kl}\delta(\sigma-\sigma')~, \qquad
  \{\psi_k(\tau,\sigma),\psi_l(\tau,\sigma')\}
  =\{\dot\psi_k(\tau,\sigma),\dot\psi_l(\tau,\sigma')\}=0~,
\end{eqnarray}
where the indices $k,l=1,2$ stand for the components of the free field 
$\psi$. Furthermore, the theory possesses extended conformal invariance, whose
generators reduce to the free field expressions after the substitution 
of the classical solution obtained from (\ref{EQ2}). In particular, the 
stress tensor generates the 
pseudo-conformal symmetry for the canonical free fields exactly the same 
manner as for the interacting fields. The virtue in writing the classical 
solution in the form (\ref{EQ2}) is that the conformal property becomes  
manifest. Since both the interacting and the free classical vertex 
functions possess the same transformation properties under the conformal 
symmetry, the four screening charges $A_aB_a$ and $A_a\circ A_{\bar a}
B_a\circ B_{\bar a}$ ($a=1,2$) must be of vanishing conformal dimensions. 
This can be verified directly from (\ref{EQ4}). 

We now turn to quantum theory. We impose the standard commutation 
relations on the normal modes
\begin{eqnarray}
  \label{EQ6}
  [Q_k,P_l]=i\delta_{kl} ~, \qquad [a_{kn}^{(+)},a_{lm}^{(+)}]
  =[a_{kn}^{(-)},a_{lm}^{(-)}]=n\delta_{kl}\delta_{n+m,0} ~.
\end{eqnarray}
Then the procedure to achieve exact operator solution for the 
interacting fields goes as follows: We first construct the exponential 
operator ${\rm e}^{\nu\lambda^a\cdot\varphi}$ with $\nu$ an arbitrary 
parameter by assuming it to be a product of a free field vertex operator 
and a power series of the screening charges with arbitrary coefficients 
depending on the zero-mode momenta $P$. These coefficients will be 
determined up to a constant order by order in the cosmological 
constant $\mu^2$ by the requirement of locality. The Toda field will 
be obtained as the derivatives of ${\rm e}^{\nu\lambda^a\cdot\varphi}$ 
with respect to $\nu$ at $\nu=0$. We then impose the field 
equations at the first order in $\mu^2$. This determine 
the unknown constant. We finally establish the full equations of motion 
and the canonical commutation 
relations. In the present work the analysis is carried out to the 
fourth order in $\mu^2$ for the general exponental operator and to 
some yet higher orders  for special cases with $\nu$ being negative 
integers. This leads us to a conjecture of the exact solution. 

To write down the Toda exponential we must find appropriate quantum 
definition for the screening charges. Since they must be $2\pi$ periodic 
in $\sigma$ and be of vanishing conformal weight, we define 
\begin{eqnarray}
  \label{EQ7}
  {\cal Y}_a(x)&=&\int_0^{2\pi}dy^+dy^-:{\cal E}_{\alpha^a}(x^+-y^+)
  {\cal E}_{\alpha^a}(x^--y^-)V_{\eta\alpha^a}(y): ~,\nonumber\\
  {\cal Y}_{a\bar a}(x)&=&\int_0^{2\pi}dy^+dy^-\int_0^{2\pi}dz^+dz^-
  :{\cal E}_{\alpha^a}(x^+-y^+){\cal E}_{\alpha^a}(x^--y^-)
  V_{\eta\alpha^a}(y): \\
  && \times:{\cal E}_{\alpha^{\bar a}}(x^+-y^+)
  {\cal E}_{\alpha^{\bar a}}(x^--y^-)
  {\cal E}_{\alpha^{\bar a}}(y^+-z^+){\cal E}_{\alpha^{\bar a}}(y^--z^-)
  V_{\eta\alpha^{\bar a}}(z): ~,\nonumber
\end{eqnarray}
where $V_\beta(x)\equiv:{\rm e}^{\beta\cdot\psi(x)}:$ is the free field 
vertex operator and : : stands for free field normal ordering for 
oscillators and symmetric ordering for zero-mode operators, i.e., 
$:{\rm e}^{\beta\cdot Q}f(P):={\rm e}^{\frac{1}{2}\beta\cdot Q}f(P)
{\rm e}^{\frac{1}{2}\beta\cdot Q}$. We have rescaled $\psi$ and $P$ 
appearing in ${\cal E}_{\alpha^b}(x)$ by a parameter $\eta$. It is fixed by requiring 
that the vertex operators 
$V_{\eta\alpha^a}$ be primary fields of weight (1,1). For 
later convenience we summarize here some notation: 
\begin{eqnarray}
  \label{EQ8}
  \varpi\equiv-\frac{iP}{\gamma\eta}~, \qquad
  \varpi^a\equiv\alpha^a\cdot\varpi ~, \qquad
  g\equiv\frac{\gamma^2\eta^2}{8\pi}~, \qquad
  q\equiv{\rm e}^{2\pi ig}~.
\end{eqnarray}
Of these $g$ plays the role of effective Planck constant and $q$ turns 
out to be the quantum deformation parameter. In terms of these variables 
the screening charges (\ref{EQ7}) can be cast into the following form 
\begin{eqnarray}
  \label{EQ7-2}
  {\cal Y}_a(x)&=&\int_0^{2\pi}dy^+dy^-
  q^{(\varpi^a+1)\varepsilon(x,y)}V_{\eta\alpha^a}(y) ~,\nonumber\\
  {\cal Y}_{a\bar a}(x)&=&\int_0^{2\pi}dy^+dy^-\int_0^{2\pi}dz^+dz^-
  q^{(\varpi^a+\varpi^{\bar a}+1)\varepsilon(x,y)
    +\varpi^{\bar a}\varepsilon(y,z)}V_{\eta\alpha^a}(y)
  V_{\eta\alpha^{\bar a}}(z) ~,
\end{eqnarray}
where we have introduced $\varepsilon(x,y)\equiv\epsilon(x^+-y^+)
+\epsilon(x^--y^-)$. 

The screening charges (\ref{EQ7}) are hermitian for the 
standard assignment of hermiticity for the normal modes. More impotantly,
one can verify the mutual commutativity of $V_{\nu\eta\lambda^a}(x)$, 
${\cal Y}_a(x)$ and ${\cal Y}_{a\bar a}(x)$. We thus expand the Toda 
exponential as 
\begin{eqnarray}
  \label{EQ9}
  {\rm e}^{\nu\lambda^a\cdot\varphi(x)}=V_{\nu\eta\lambda^a}(x)
  \sum_{n,m=0}^\infty\Biggl(\frac{\mu^2}{4}\Biggr)^{n+2m}
  C^a_{nm}(\varpi;\nu){\cal Y}^n_a(x)
  {\cal Y}^m_{a\bar a}(x) ~,
\end{eqnarray}
where the coefficients $C^a_{nm}$ may depend on the zero-mode momenta 
without affecting the conformality and we assume $C^a_{00}(\varpi,\nu)=1$. 
Note that there arises no ordering ambiguity due to the commutativity of 
operators metioned above. This greatly simplifies the following analysis. 

Strictly speaking, the operator product $V_{\nu\eta\lambda^a}{\cal Y}_a^n
{\cal Y}^m_{a\bar a}$ becomes ill-defined on the physically 
interesting region of $\nu,g$ for sufficiently large $n,m$ 
\cite{ow,cgs,kn}. We 
may, however, consider it as an analytic continuation from the region 
where it is well-defined \cite{cgs}. 

To determine the coefficients $C^a_{nm}$, we require the locality conditions 
$[{\rm e}^{\kappa\lambda^a\cdot\varphi(\tau,\sigma)},
{\rm e}^{\nu\lambda^b\cdot\varphi(\tau,\sigma')}]=0$ for $\sigma\neq\sigma'$. 
These lead to the following constraints 
\begin{eqnarray}
  \label{EQ10}
  \sum_{{n+r+2(m+s)=J}\atop {n+m+r+s=K}}[C^a_{nm}(\varpi+\kappa\lambda^a;
  \kappa)C^a_{rs}(\varpi+(\kappa+\nu)\lambda^a+(n+m)\alpha^a
  +m\alpha^{\bar a};\nu)I^a_{nm}{}^{;a}_{;rs}
  (\kappa,\nu;x,x') 
  \nonumber\\
  -C^a_{rs}(\varpi+\nu\lambda^a;\nu)C^a_{nm}(\varpi+
  (\kappa+\nu)\lambda^a+(r+s)\alpha^a+s\alpha^{\bar a};\kappa)
  I^a_{rs}{}^{;a}_{;nm}(\nu,\kappa;x',x)]=0 ~,\\
  \label{EQ10-2}
  \sum_{{n+r+2(m+s)=J}\atop {n+m+s=K}}[C^a_{nm}(\varpi+\kappa\lambda^a;
  \kappa)C^{\bar a}_{rs}(\varpi+\kappa\lambda^a+\nu\lambda^{\bar a}
  +(n+m)\alpha^a+m\alpha^{\bar a};\nu)I^a_{nm}{}^{;{\bar a}}_{;rs}
  (\kappa,\nu;x,x')
  \nonumber\\
  -C^{\bar a}_{rs}(\varpi+\nu\lambda^{\bar a};\nu)C^a_{nm}(\varpi+
  \kappa\lambda^a+\nu\lambda^{\bar a}+(r+s)\alpha^{\bar a}+s\alpha^a;\kappa)
  I^{\bar a}_{rs}{}^{;a}_{;nm}(\nu,\kappa;x',x)]=0 
\end{eqnarray}
with $x^\pm=\tau\pm\sigma$ and $x'{}^\pm=\tau\pm\sigma'$. We have introduced 
\begin{eqnarray}
  \label{EQ11}
  I^a_{nm}{}^{;b}_{;rs}(\kappa,\nu;x,x')\equiv
  V_{\kappa\eta\lambda^a}(x){\cal Y}^n_a(x)
  {\cal Y}^m_{a\bar a}(x)V_{\nu\eta\lambda^b}(x')
  {\cal Y}^r_b(x'){\cal Y}^s_{b\bar b}(x') ~.
\end{eqnarray}
The sum should be taken over nonnegative integers $n,m,r,s$ for 
given $J$, the order of $\mu^2$, and $K$, the number of 
$V_{\eta\alpha^a}$ contained in the operator products, with
$J\ge K$. Due to the symmetry $a\leftrightarrow \bar a$ we 
have only to consider the case $a=1$. 

At the lowest order, the locality 
conditions for $J=K=0$ are trivially satisfied since the free field 
vertex operators commute at equal time. To illustrate how (\ref{EQ10}) 
and (\ref{EQ10-2}) work,
we consider the case $J=K=1$. The constraint (\ref{EQ10}) can be 
cast into the following form
\begin{eqnarray}
  \label{EQ11-2}
  &&\int_0^{2\pi}dy^+dy^-[q^{(\varpi^a+\kappa+1)\varepsilon(x,y)
    +\nu\varepsilon(x',y)}
  C^a_{10}(\varpi+\kappa\lambda^a;\kappa)
  +q^{(\varpi^a+\kappa+1)\varepsilon(x',y)}
  C^a_{10}(\varpi+(\kappa+\nu)\lambda^a;\nu)\nonumber\\
  &&\hskip 2.0cm -q^{(\varpi^a+\kappa+\nu+1)\varepsilon(x,y)}
  C^a_{10}(\varpi+(\kappa+\nu)\lambda^a;\kappa)
  -q^{(\varpi^a+\nu+1)\varepsilon(x',y)+\kappa\varepsilon(x,y)}
  C^a_{10}(\varpi+\nu\lambda^a;\nu)]\nonumber\\
  &&\hskip 2.5cm \times V_{\kappa\eta\lambda^a}(x)V_{\nu\eta\lambda^a}(x')
  V_{\eta\alpha^a}(y)=0 ~.
\end{eqnarray}
Such  relation holds true only when the 
integrand identically vanishes. After some algebraic manipulations 
we obtain 
\begin{eqnarray}
  \label{EQ12}
  &&C^a_{10}(\varpi-\nu\lambda^a;\kappa)
  +q^{-(\varpi^a-\nu+1)\varepsilon(x,y,x')}C^a_{10}(\varpi;\nu) 
  \nonumber\\
  &&-q^{\nu\varepsilon(x,y,x')}C^a_{10}(\varpi;\kappa)
  -q^{-(\varpi^a-\kappa-\nu+1)\varepsilon(x,y,x')}
  C^a_{10}(\varpi-\kappa\lambda^a;\nu)=0~,
\end{eqnarray}
where we have introduced $\varepsilon(x,y,z)\equiv \varepsilon(x,y)+
\varepsilon(y,z)$. 
Since $\varepsilon(x,y,x')$ takes the values $0,\pm 2$ and cannot be 
equal to $\pm4$, this is equivalent to the three independent relations
\begin{eqnarray}
  \label{EQ13}
  C^a_{10}(\varpi-\nu\lambda^a;\kappa)+C^a_{10}(\varpi;\nu)
  -C^a_{10}(\varpi;\kappa)-C^a_{10}(\varpi-\kappa\lambda^a;\nu)=0 ~,
  \nonumber\\
  C^a_{10}(\varpi-\nu\lambda^a;\kappa)
  +q^{2(\varpi^a-\nu+1)}C^a_{10}(\varpi;\nu)
  -q^{-2\nu}C^a_{10}(\varpi;\kappa)
  -q^{2(\varpi^a-\kappa-\nu+1)}C^a_{10}(\varpi-\kappa\lambda^a;\nu)=0 ~, \\
  C^a_{10}(\varpi-\nu\lambda^a;\kappa)
  +q^{-2(\varpi^a-\nu+1)}C^a_{10}(\varpi;\nu)
  -q^{2\nu}C^a_{10}(\varpi;\kappa)
  -q^{-2(\varpi^a-\kappa-\nu+1)}C^a_{10}(\varpi-\kappa\lambda^a;\nu)=0 ~.
  \nonumber
\end{eqnarray}
While the constraint (\ref{EQ10-2}) leads to 
\begin{eqnarray}
  \label{EQ14}
  C^a_{10}(\varpi-\nu\lambda^{\bar a};\kappa)=C^a_{10}(\varpi;\kappa)~.
\end{eqnarray}
This implies that $C^a_{10}(\varpi;\kappa)$ is independent of 
$\varpi^{\bar a}$ as is expected since the lowest order analysis 
is essentially the same with that of Liouville theory. The solution 
to (\ref{EQ13}) has been argued in detail in refs. \cite{ow,kn}. 
We simply give the result for the generic value of $g$ 
\begin{eqnarray}
  \label{EQ15}
  C^a_{10}(\varpi;\nu)=\frac{c_0[\nu]}{[\varpi^a+1][\varpi^a-\nu+1]}~,
\end{eqnarray}
where we have introduced $q$-numbers $[x]\equiv\displaystyle{
\frac{q^x-q^{-x}}{q-q^{-1}}}$. The arbitrary constant $c_0$ can be fixed 
by imposing the field equations at the first order in 
$\mu^2$, where we can replace the Toda fields 
in the Toda potential by $\eta\psi$. This leads to  
\begin{eqnarray}
  \label{EQ16}
  c_0=(8\pi g\sin2\pi g)^{-1}~.
\end{eqnarray}
We generalize all this to $C^a_{n0}(\varpi;\kappa)$ for arbitrary $n$ 
since (\ref{EQ10}) for $J=K$ are equivalent to the locality conditions 
in the Liouville case. One can obtain these coefficients from the 
results of ref. \cite{fit96}.

At the second order in $\mu^2$, we have only to determine $C^a_{01}$. 
This can be achieved by considering the 
constraint (\ref{EQ10-2}) for $J=2$ and $K=1$. By the analysis 
similar to the first order case we obtain the condition
\begin{eqnarray}
  \label{EQ17}
  && q^{(\varpi^{\bar a}-\nu)\varepsilon(x,y_1,y_2,x')}
  C^a_{01}(\varpi-\nu\lambda^{\bar a};\kappa)
  +q^{-(\varpi^a+1)\varepsilon(x,y_1,y_2,x')}
  C^{\bar a}_{01}(\varpi;\nu) \nonumber\\
  &&\hskip .25cm -q^{\varpi^{\bar a}\varepsilon(x,y_1,y_2,x')}
  C^a_{01}(\varpi;\kappa)
  -q^{-(\varpi^a-\kappa+1)\varepsilon(x,y_1,y_2,x')}
  C^{\bar a}_{01}(\varpi-\kappa\lambda^a;\nu) \nonumber\\
  &&\hskip .5cm =-C^a_{01}(\varpi-\nu\lambda^{\bar a};\kappa)
  C^{\bar a}_{10}(\varpi+\alpha^a;\nu)+q^{-\varepsilon(x,y_1,y_2,x')}
  C^a_{10}(\varpi+\alpha^{\bar a};\kappa)
  C^{\bar a}_{10}(\varpi-\kappa\lambda^a;\nu) ~,
\end{eqnarray}
where $\varepsilon(x,y,z,w)\equiv\varepsilon(x,y)
+\varepsilon(y,z)+\varepsilon(z,w)$. Since 
$\varepsilon(x,y_1,y_2,x')$ takes only the values $0,\pm2,\pm4$, 
(\ref{EQ17}) implies five 
constraints. They can be solved with respect to $C^a_{01}(\varpi;\kappa)$. 
We thus obtain the second order coefficients
\begin{eqnarray}
  \label{EQ18}
  C^a_{20}(\varpi;\nu)&=&\frac{c_0^2[\nu][\nu+1]}{[2][\varpi^a+2][\varpi^a+3]
    [\varpi^a-\nu+1][\varpi^a-\nu+2]} ~,\nonumber\\
  C^a_{01}(\varpi;\nu)&=&-\frac{c_0^2[\nu]}{[\varpi^a+\varpi^{\bar a}+1]
    [\varpi^a+\varpi^{\bar a}-\nu+1][\varpi^{\bar a}][\varpi^{\bar a}+1]} ~.
\end{eqnarray}

The analysis can be carried out for higher orders. The complication of 
the constraints (\ref{EQ10}) and (\ref{EQ10-2}), however, grows rapidly 
with the order of $\mu^2$. We only quote here the results of the third and 
the fourth order cases. 

\vskip .2cm\noindent
Third order coefficients: 
\begin{eqnarray}
  \label{EQ19}
  C^a_{30}(\varpi;\nu)&=&\frac{c_0^3[\nu][\nu+1][\nu+2]}{[2][3][\varpi^a+3]
    [\varpi^a+4][\varpi^a+5]
    [\varpi^a-\nu+1][\varpi^a-\nu+2][\varpi^a-\nu+3]} ~,\nonumber\\
  C^a_{11}(\varpi;\nu)&=&-\frac{c_0^3[\nu][\nu+1]}{[\varpi^a+2][\varpi^a-\nu+1]
    [\varpi^a+\varpi^{\bar a}+2][\varpi^a+\varpi^{\bar a}-\nu+1]
    [\varpi^{\bar a}-1][\varpi^{\bar a}+1]} ~.
\end{eqnarray}

\vskip .2cm\noindent
Fourth order coefficients:
\begin{eqnarray}
  \label{EQ20}
  C^a_{40}(\varpi;\nu)&=&c_0^4\frac{[\nu][\nu+1][\nu+2][\nu+3]}{[2][3][4]}
    \frac{1}{[\varpi^a+4][\varpi^a+5][\varpi^a+6][\varpi^a+7]} \nonumber\\
    &&\times\frac{1}{[\varpi^a-\nu+1][\varpi^a-\nu+2][\varpi^a-\nu+3]
      [\varpi^a-\nu++4]} ~,
  \nonumber\\
  C^a_{21}(\varpi;\nu)&=&-c_0^4\frac{[\nu][\nu+1][\nu+2]}{[2]}
  \frac{1}{[\varpi^a+3][\varpi^a+4][\varpi^a-\nu+1][\varpi^a-\nu+2]}
  \nonumber\\
  &&\times\frac{1}{[\varpi^a+\varpi^{\bar a}+3]
    [\varpi^a+\varpi^{\bar a}-\nu+1]}
  \frac{1}{[\varpi^{\bar a}-2][\varpi^{\bar a}+1]} ~, \nonumber \\
  C^a_{02}(\varpi;\nu)&=&c_0^4\frac{[\nu][\nu+1]}{[2]} 
  \frac{1}{[\varpi^a+\varpi^{\bar a}+2][\varpi^a+\varpi^{\bar a}+3]
    [\varpi^a+\varpi^{\bar a}-\nu+1][\varpi^a+\varpi^{\bar a}-\nu+2]}
  \nonumber\\
  &&\times\frac{1}{[\varpi^{\bar a}][\varpi^{\bar a}+1]^2
    [\varpi^{\bar a}+2]} ~. 
\end{eqnarray}

These results suffice to infer the general form of $C^a_{nm}$ for 
arbitrary $n,m$. We thus arrive at the conjecture 
for the coefficients
\begin{eqnarray}
  \label{EQ22}
  C^a_{nm}(\varpi;\nu)&=&(-1)^mc_0^{n+2m}
  \frac{\Gamma_q[\nu+n+m]}{[n]![m]!\Gamma_q[\nu]}
  \frac{\Gamma_q[\varpi^a+\varpi^{\bar a}+n+m]
    \Gamma_q[\varpi^a+\varpi^{\bar a}-\nu+1]}{
    \Gamma_q[\varpi^a+\varpi^{\bar a}+n+2m]
    \Gamma_q[\varpi^a+\varpi^{\bar a}-\nu+m+1]}
  \nonumber\\
  &&\times\frac{\Gamma_q[\varpi^a+n+m]\Gamma_q[\varpi^a-\nu+1]}{
    \Gamma_q[\varpi^a+2n+m]\Gamma_q[\varpi^a-\nu+n+1]}
  \frac{\Gamma_q[\varpi^{\bar a}-n]\Gamma_q[\varpi^{\bar a}+1]}{
    \Gamma_q[\varpi^{\bar a}-n+m]\Gamma_q[\varpi^{\bar a}+m+1]}~,~~~~
\end{eqnarray}
where $\Gamma_q[x]$ is the $q$-deformed $\Gamma$-function defined by 
$\Gamma_q[x+1]=[x]\Gamma_q[x]$ with $\Gamma_q[1]=1$, and $[n]!\equiv
\Gamma_q[n+1]$ stands for the $q$-factorial. With (\ref{EQ16}) and 
(\ref{EQ22}), (\ref{EQ9}) represents the $q$-deformation of the 
classical Toda exponential. In fact one can show that in the 
limit $g\rightarrow0$ with $8\pi g\varpi\rightarrow \gamma P$ kept 
finite our operator expressions reduce to the classical solution 
(\ref{EQ2}). This is the central result of the present work. 

So far our analysis has been restricted to the Toda exponentials
associated with the fundamental weights. We can, however, define 
arbitrary exponential operators ${\rm e}^{\beta\cdot\varphi}$ in 
terms of them as a composite operator:
\begin{eqnarray}
  \label{EQ23}
  {\rm e}^{\beta\cdot\varphi}&=& \Biggl(\sum_{n,m=0}^\infty 
  \Biggl(\frac{\mu^2}{4}\Biggr)^{n+2m}C^a_{nm}
  (\varpi+\alpha^a\cdot\beta\lambda^a;\alpha^a\cdot\beta)
  {\cal Y}^n_a{\cal Y}^m_{a\bar a}\Biggr)V_{\eta\beta} \nonumber\\
  &&\times
  \Biggl(\sum_{r,s=0}^\infty \Biggl(\frac{\mu^2}{4}\Biggr)^{r+2s}
  C^{\bar a}_{rs}
  (\varpi;\alpha^{\bar a}\cdot\beta){\cal Y}^r_{\bar a}{\cal Y}^s_{\bar aa}
  \Biggr) ~,
\end{eqnarray}
where $\beta$ is an arbitrary vector in the $A_2$-root space. In particular 
we need such operators for 
$\beta=\alpha^a$ to verify the field equations. 

The Toda exponential (\ref{EQ9}) reduces to a finite polynomial 
in the screening charges when $\nu$ is a negative integer as in the 
Liouville case. 
The composition rule (\ref{EQ23}) can be used to construct such operators 
as the $(-\nu)$-th power of ${\rm e}^{-\lambda^a\cdot\varphi}$, which is 
given exactly in our analysis. Since the latter operator satisfies the 
locality, any composite of such operators is necessarily local. It can 
be shown that the coefficients $C^a_{nm}(\varpi;\nu)$ can be obtained 
recursively from 
\begin{eqnarray}
  \label{EQ21}
  &&\sum_{n,m}\Biggl(\frac{\mu^2}{4}\Biggr)^{n+2m}C^a_{nm}(\varpi;\nu)
  {\cal Y}_a^n{\cal Y}^m_{a\bar a}\nonumber\\
  &&\hskip .5cm=\prod_{k=0}^{-\nu-1}(1+\frac{\mu^2}{4}
  C^a_{10}(\varpi+(k-\nu -1)\lambda^a;-1)
  {\cal Y}_a+\Biggl(\frac{\mu^2}{4}\Biggr)^2
  C^a_{01}(\varpi+(k-\nu-1)\lambda^a;-1){\cal Y}_{a\bar a})~.~~~~~
\end{eqnarray}
This can be used to check the validity of (\ref{EQ22}). We have worked out 
the fifth order coefficients for $\nu$ being some negative integers. They 
are consistent with (\ref{EQ22}).

To complete our construction of the exact operator solution, we must
verify the field equations and the canonical commutaion 
relations for the interacting Toda field obtained from $\lambda^a
\cdot\varphi\equiv\displaystyle{\frac{d}{d\nu}{\rm e}^{\nu\lambda^a
\cdot\varphi}\Bigg|_{\nu=0}}$. This can be carried out in exactly the 
same way for the locality analysis. We have confirmed that the Toda 
fields satisfy the operatorial field equations and the canonical 
commutaion relations
\begin{eqnarray}
  \label{EQ24}
  [\varphi_j(\tau,\sigma),\dot\varphi_k(\tau,\sigma')]
  =[\eta\psi_j(\tau,\sigma),\eta\dot\psi_k(\tau,\sigma')] 
\end{eqnarray}
to the fourth order in $\mu^2$. Thus the operatorial mapping 
$\eta\psi\rightarrow\varphi$ can be considered as canonical. 

In summary we have investigated quantum $A_2$-Toda field theory 
based on the method of quantizing canonical free fields. We have 
obtained exact operator solution satisfying the fundamental requirements 
of quantum theory to fourth order of the cosmological constant. This 
lead us to the conjecture for the full order expression. Though the 
analysis have been restricted to $A_2$ case, it can be extended 
to other Toda theories of higher ranks. Our results also suggests that 
the algebraic approach developed for Liouville 
theory \cite{cgs,fit96} can be extended to Toda field theories. We 
will argue the issue elsewhere. 

\eject

\end{document}